\documentclass[showpacs,aps,graphicx,twocolumn]{revtex4}
\usepackage{graphicx}

\begin{document}

\title{Multiparty-controlled teleportation of an arbitrary $m$-qudit state with pure entangled quantum channel
\footnote{Published in \emph{J. Phys. A: Math. Theor.} \textbf{40}
(2007) 13121-13130}}
\author{Ping Zhou,$^{1,2,3}$ Xi-Han Li,$^{1,2,3}$ Fu-Guo Deng,$^{1,2,3}$\footnote{ E-mail: fgdeng@bnu.edu.cn} and Hong-Yu Zhou$^{1,2,3}$ }
\address{$^{1}$ Key Laboratory of Beam Technology and Material
Modification of Ministry of Education, Beijing Normal University,
Beijing 100875, People's Republic of China\\
$^{2}$ Institute of Low Energy Nuclear Physics, and Department of
Material Science and Engineering, Beijing Normal University,
Beijing 100875, People's Republic of China\\
$^{3}$ Beijing Radiation Center, Beijing 100875, People's Republic
of China}
\date{\today }

\begin{abstract}
We present a general scheme for multiparty-controlled teleportation
of an arbitrary $m$-qudit ($d$-dimensional quantum system) state by
using non-maximally entangled states as the quantum channel. The
sender performs $m$ generalized Bell-state measurements on her 2$m$
particles, the controllers take some single-particle measurements
with the measuring basis $X_{d}$ and the receiver only need to
introduce one auxiliary two-level particle to extract quantum
information probabilistically with the fidelity unit if he
cooperates with all the controllers. All the parties can use some
decoy photons to set up their quantum channel securely, which will
forbid some a dishonest party to eavesdrop freely. This scheme is
optimal as the probability that the receiver obtains the originally
unknown $m$-qudit state equals to the entanglement of the quantum
channel.
\end{abstract}
 \maketitle

\section{introduction}

The principle of quantum mechanics provides some novel ways for
quantum information transmission, such as quantum key distribution
\cite{BB84,Gisin,QKD,QKD2}, quantum secret sharing
\cite{HBB99,KKI,longqss,dengsinglephoton,improve}, quantum secure
direct communication \cite{QSDC,QSDC2}, deterministic secure quantum
communication \cite{DSQC,DSQC2}, quantum secret report
\cite{dengreport}, quantum secret conference \cite{conference},
quantum dialogue \cite{dialogue}, quantum teleportation
\cite{teleportation}, and so on. Quantum teleportaiton, a unique
thing in quantum mechanics, supplies a way for two parties to
teleport an unknown quantum state $\vert\chi\rangle=a\vert 0\rangle
+ b\vert 1\rangle$, exploiting the nonlocal correlation of an
Einstein-Podolsky-Rosen (EPR) state shared in advance. For this
task, the sender performs a Bell-state measurement on the unknown
quantum system $\chi$ and one of the EPR particles, and the receiver
takes a unitary operation on the remaining EPR particle, according
to the information of the Bell-state measurement. Since Bennet
\emph{et al.} \cite{teleportation} first discovered that the
information of an unknown qubit  $\vert\chi\rangle$ can be
disassembled into some pieces and then reconstructed with classical
information and quantum correlations, researchers have devoted much
interest to quantum teleportation.  On one hand, several experiments
have demonstrated the teleportation of a single qubit with entangled
photons and ions \cite{TE1,TE2,TE3,TE4,TE5,TE52}.  On the other
hand, a great number of theoretical schemes for teleporting an
unknown state, especially an $N$-particle entangled state, have been
proposed with different quantum channels
\cite{TT1,TT2,ShiGuo,LuGuo,Lee,yannmb,YanFLteleportation1,probabilistic1,qudit1,qudit2,yanpla,YangGuo,Rigolin,dengprac}.

Recently, the controlled teleportation for single-qubit or $m$-qubit
state have been studied by some groups.  The basic idea of a
controlled teleportation scheme \cite{first,yang,dengghz,lixhcpl} is
to let an unknown quantum state be recovered by a remote receiver
only when he cooperates with the controllers. It is similar to
another branch of quantum communication, quantum state sharing
(QSTS)
\cite{peng,dengepr,lixhjpb,zhangzjepjd,dengepjd,zhangscp,GQSTS},
whose task is to let several receivers share an unknown quantum
state with cooperations. Essentially one receiver can reconstruct
the originally unknown state with the help of others. In principle,
almost all the QSTS schemes
\cite{peng,dengepr,lixhjpb,zhangzjepjd,dengepjd,zhangscp,GQSTS} can
be used for controlled teleportation with or without a little
modification, and vice versa \cite{dengghz,dengepr,lixhjpb}. In
1999, Karlsson and Bourennane proposed the first controlled
teleportation protocol with three-qubit Greenberger-Horne-Zeilinger
(GHZ) state for teleporting a single-qubit state \cite{first}. In
2004, Yang \emph{et al.} \cite{yang} presented a multiparty
controlled teleportation protocol to teleport multi-qubit quantum
information. In  2005, Deng \emph{et al.} \cite{dengghz} introduced
a symmetric multiparty controlled teleportation scheme for an
arbitrary two-particle entanglement state. Moreover, they presented
another scheme for sharing an arbitrary two-particle state with EPR
pairs and GHZ-state measurements \cite{dengepr} or Bell-state
measurements \cite{dengepjd}. Both those two QSTS schemes
\cite{dengghz,dengepr} can be used for controlled teleportation
directly without any modification. Also, Zhang, Jin and Zhang
\cite{zhangscp} presented a scheme for sharing an arbitrary
two-particle state based on entanglement swapping. Zhang \emph{et
al.} \cite{zhangzjepjd} proposed a multiparty QSTS scheme for
sharing an unknown single-qubit state with photon pairs and a
controlled teleportation scheme by using quantum secret sharing of
classical message for teleporting arbitrary $m$-qubit quantum
information. Recently, Li \emph{et al.} \cite{lixhjpb} proposed an
efficient symmetric multiparty QSTS protocol for sharing an
arbitrary $m$-qubit state. All those three QSTS schemes, in
principle, are equivalent to a secure scheme for teleportation with
some controllers.

Although  there are some schemes for controlled teleportation
\cite{first,yang,dengghz,lixhcpl} or QSTS
\cite{peng,dengepr,lixhjpb,zhangzjepjd,dengepjd,zhangscp}, all of
them are based on a maximally entangled quantum channel, not a pure
entangled one. A practical quantum signal source often produces a
pure entangled state because of its unsymmetry to some extent. In
this work, we will give a general form for controlled teleportation
of an arbitrary $m$-qudit ($d$-level quantum system) state via  the
control of $n$ controllers by using $d$-dimensional pure entangled
states as the quantum channel, following some ideas in Ref.
\cite{lixhjpb}. Except for the sender Alice, each of the controllers
needs only to take $m$ single-particle measurements on his
particles, and the receiver can probabilistically reconstruct the
unknown $m$-qudit state with an auxiliary qubit (two-level particle)
and $m$ unitary operations if she cooperates with all the
controllers. This scheme for controlled teleportation of $m$-qudit
state is optimal as the probability that the receiver obtains the
originally unknown $m$-qudit state with the fidelity unit equals to
the entanglement of the quantum channel.

\section{controlled teleportation of an arbitrary single-particle qudit with pure entangled quantum channel}

The generalized Bell states (GBS) of $d$-dimensional quantum systems
(the analogue of Bell state for spin- $\frac{1}{2}$ particles) are
\cite{teleportation}
\begin{eqnarray}
\vert \psi_{rs}\rangle=\frac{1}{\sqrt{d}}\sum_{j=0}^{d-1}
e^{\frac{2\pi i }{d}j r}\vert j\rangle\vert j\oplus s\rangle
\end{eqnarray}
where $r,s=0,1,\cdots, d-1$ are used to label the $d^2$ orthogonal
GBS. $\vert 0\rangle$, $\vert 1\rangle$, $\cdots$, and $\vert
d-1\rangle$ are the $d$ eigenvectors of the measuring basis (MB)
$Z_{d}$, and $j\oplus s$ means $j+s$ mod $d$. The $ d^{2} $ unitary
operations $U_{uv}$ $(u,v=0,1,\cdots, d-1)$ can transfer one of the
Bell state into each other.
\begin{eqnarray}
U_{uv}=\sum_{j=0}^{d-1}e^{\frac{2\pi i }{d} u j} \left\vert j\oplus
v\right\rangle \left\langle j\right\vert.
\end{eqnarray}
Another unbiased basis $X_{d}$ which has $d$ eigenvectors can be
written as $\{\vert 0\rangle_{x}, \cdots, \vert r\rangle_x, \cdots,
\vert d-1\rangle_{x}\}$ \cite{DSQC,chenp}.
\begin{eqnarray}
\vert r\rangle_{x}=\frac{1}{\sqrt{d}}\sum_{j=0}^{d-1}e^{\frac{2\pi i
}{d}j r}\vert j\rangle
\end{eqnarray}
where $r\in\{0,1,\cdots,d-1\}$. The two unbiased bases have the
relation $|\langle k|r\rangle_x|^2=\frac{1}{d}$. Here $\vert
k\rangle$ is an eigenvector of the MB $Z_d$ and $\vert r\rangle_x$
is an eigenvector of the MB $X_d$.

Now, let us describe the principle of our controlled teleportation
of an arbitrary $m$-qudit state with $m$ pure entangled states. For
presenting the principle of our scheme clearly, we first consider
the case to teleport an unknown single-particle qudit state and then
generalized it to the case with an arbitrary $m$-particle qudit
state.

Suppose the originally unknown single-particle qudit state
teleported is
\begin{eqnarray}
\left\vert \chi \right\rangle_{\chi_{0}}=\beta_{0}\left\vert
0\right\rangle +\beta_{1}\left\vert 1\right\rangle +\cdots +
\beta_{d-1}\left\vert d-1\right\rangle,
\end{eqnarray}
here
\begin{eqnarray}
\vert \beta_{0}\vert^2 + \vert \beta_{1}\vert^2 + \cdots + \vert
\beta_{d-1}\vert^2=1.
\end{eqnarray}
The pure entangled $(n+2)$-particle state used for setting up the
quantum channel is
\begin{eqnarray}
\left\vert\Phi\right\rangle_{a_0a_1\cdots
a_{n+1}}=c_{0}\prod_{k=0}^{n+1}\left\vert 0\right\rangle_{a_k}
+\cdots+\prod_{k'=0}^{n+1}c_{d-1}\left\vert
d-1\right\rangle_{a_{k'}},
\end{eqnarray}
where $a_k (k=0,1,\cdots,n+1)$ are the $n+2$ particles in the pure
entangled state $\vert \Phi\rangle$,  and
\begin{eqnarray}
\frac{1}{d}\sum_{j=0}^{d-1} \vert c_j\vert^2=1.
\end{eqnarray}

Similar to the controlled teleportation of qubits
\cite{dengghz,lixhjpb}, Alice should first set up a pure entangled
quantum channel with the controllers, say Bob$_q$ ($q=1,2,\cdots,n$)
and the receiver, say Charlie. The way for sharing a sequence of
pure entangled $(n+2)$-particle qubit states has been discussed in
Ref. \cite{DSQC}. In detail, Alice prepares a sequence of pure
entangled states $\vert \Phi\rangle_{a_0a_1\cdots a_{n+1}}$, and
divided them into $n+2$ particle sequences, say $S_k$($k=0,1,
\cdots,n+1$). That is, Alice picks up the particle $a_k$ in each
pure entangled state $\vert \Phi\rangle_{a_0a_1\cdots a_{n+1}}$ to
make up the particle sequence $S_k$, as the same way as Refs.
\cite{QSDC,QSDC2,dengghz,lixhjpb}. To prevent  a potentially
dishonest controller from stealing the information freely or the
receiver from recovering the unknown state without the control of
the controllers \cite{dengattack}, Alice has to replace some
particles in the sequence $S_k$ with her decoy photons
\cite{decoy,decoy2} before she sends the sequence $S_k$ to a
controller, say Bob$_k$ (or the receiver Charlie if $k=n+1$). The
decoy photons can be prepared by measuring and manipulating some
particles in pure entangled states \cite{DSQC}. For instance, Alice
measures the particle $a_0$ in the state
$\left\vert\Phi\right\rangle_{a_0a_1\cdots a_{n+1}}$ with the MB
$Z_d$, and then obtains the state of all the other particles $\vert
r\rangle$ if that of the particle $a_0$ is $\vert r\rangle$. Alice
can manipulate the particle $a_k$ with unitary operations
$\{U'_{uv}=\vert u\rangle \langle v\vert \}$ and high-dimensional
Hadamard operation $H_d$ \cite{decoy2,DSQC,chenp}.
\begin{eqnarray}
H_d  =\frac{1}{\sqrt{d}} \left( {\begin{array}{*{20}c}
   1 & 1 &  \cdots  & 1  \\
   1 & {e^{2\pi i/d} } &  \cdots  & {e^{(d-1)2\pi i/d} }  \\
   1 & {e^{4\pi i/d} } &  \cdots  & {e^{(d-1)4\pi i/d}  }\\
    \vdots  &  \vdots  &  \cdots  & \vdots  \\
   1 & {e^{2(d-1)\pi i/d} } &  \cdots  & {e^{(d-1)2(d-1)\pi i/d} }  \\
\end{array}} \right)\label{HD}.
\end{eqnarray}
That is, Alice can prepares her decoy photons randomly in one of the
2$d$ states $\{\vert 0\rangle, \vert 1\rangle, \cdots, \vert
d-1\rangle; \vert 0\rangle_x, \vert 1\rangle_x, \cdots, \vert
d-1\rangle_x\}$ without an ideal high-dimension single-photon source
\cite{DSQC,decoy2}.

After setting up the pure entangled quantum channel securely, Alice
performs a generalized Bell-state measurement on her particles
$\chi_0$ and $a_{0}$, the quantum correlation will be transferred to
the quantum system composed of the other $n+1$ particles $a_{1}$,
$a_{2}$, $\cdots$, and $a_{n+1}$. For reconstructing the original
qudit state $\vert \chi\rangle_{\chi_0}$, the  $n$ controllers
Bob$_k$ perform $X_{d}$ measurements on their particles and the
receiver can probabilistically extract the information of the
original state $\vert \chi\rangle_{\chi_0}$ by introducing one
auxiliary two-level particle. In detail, one can rewrite the state
of composite quantum system composed of all the  particles $
\chi_{0}$, $a_0$, $a_1$, $\cdots$, and $a_{n+1}$ as follows:
\begin{eqnarray}
\vert \chi\rangle_{\chi_{0}}&\otimes & \vert
\Phi\rangle_{a_{0}a_1\cdots a_{n+1}} = \left(\sum_{j=0}^{d-1}
\beta_{j}\vert j \rangle\right)_{\chi_{0}}\nonumber\\
& \otimes & \left(\sum_{j'=0}^{d-1} c_{j'}\prod_{k=0}^{n+1} \vert
j'\rangle_{a_{k}}\right)
\nonumber\\
&=& \frac{1}{\sqrt{d}}\sum_{r,s}[\vert\psi_{rs}\rangle_{\chi_{0}a_{0}}\nonumber\\
& \otimes & \sum_{j=0}^{d-1}e^{-\frac{2\pi i }{d}jr}\beta_j
c_{j\oplus s}\prod_{k=1}^{n+1}\vert j\oplus s\rangle_{a_{k}}].
\end{eqnarray}
After Alice performs the generalized Bell-state (GBS) measurement on
the particles $\chi_0$ and $a_0$, the remaining particles ($a_1,
a_2, \cdots, a_{n+1}$) collapse to the state
$\vert\varphi\rangle_{a_{1}\cdots a_{n+1}}$ (without being
normalized) if Alice gets the outcome
$\vert\psi_{rs}\rangle_{\chi_{0}a_{0}}$.
\begin{eqnarray}
\vert\varphi\rangle_{a_{1},\cdots,a_{n+1}}=\sum_{j=0}^{d-1}e^{-\frac{2\pi
i }{d}jr}\beta_j c_{j \oplus s} \prod_{k=1}^{n+1}\vert j\oplus
s\rangle_{a_{k}}.
\end{eqnarray}
To probabilistically reconstruct the original state, the controllers
Bob$_k$ perform  measurements with the MB $X_{d}$ on their particles
independently. The measurements done by all the controllers can be
expressed as $M$, similar to Refs. \cite{lixhjpb,dengepr}.
\begin{eqnarray}
M=(\langle 0\vert_{x})^{n-t_{1}-t_2-\cdots -t_{d-1}}\otimes (\langle
1\vert_x)^{t_1}\otimes \cdots\otimes(\langle
d-1\vert_{x})^{t_{d-1}}.\nonumber\\
\end{eqnarray}
Here $t_{j}$ ($j=1,2,\cdots, d-1$) represents the number of the
controllers that obtain the result $\vert j\rangle_{x}$. After
controllers perform $M$ measurement on their particles, the state of
the  particle in the hand of the receiver Charlie becomes (neglect a
whole factor $1/{d^{n/2}}$)
\begin{eqnarray}
\vert\varphi\rangle_{a_{n+1}}&=&M(\sum_{j=0}^{d-1}e^{-\frac{2\pi i
}{d}jr}\beta_j c_{j \oplus s} \prod_{k=1}^{n+1}\vert j\oplus
s\rangle_{a_{k}})\nonumber\\&=&\sum_{j=0}^{d-1}e^{-\frac{2\pi i
}{d}[jr+(j\oplus s)r']}\beta_{j}c_{j \oplus s}\vert j \oplus
s\rangle_{a_{n+1}},
\end{eqnarray}
where
\begin{eqnarray}
 r'=1\cdot t_{1}+ 2 \cdot t_2 +\cdots +(d-1)\cdot t_{d-1}.
\end{eqnarray}
That is, the state of the receiver's particle $a_{n+1}$ is
determined by the measurement results of the sender and all the
controllers. Suppose $|c_{k}|^2=min\{|c_{i}|^2,i=0,\cdots,d-1\}$.
For extracting information of the original state $\vert
\chi\rangle_{\chi_{0}}$ from $\vert\varphi\rangle_{a_{n+2}}$
probabilistically, Charlie can perform a general evolution $U_{max}$
on particle $a_{n+1}$ and an auxiliary qubit $a_{aux}$ whose
original state is $\vert 0\rangle_{aux}$. In detail, under the basis
$\{\vert 0 \rangle\vert0\rangle_{aux}, \vert
1\rangle\vert0\rangle_{aux}, \cdots, \vert
d-1\rangle\vert0\rangle_{aux},\,\,\,\, \vert 0\rangle\vert
1\rangle_{aux}, \cdots,\vert d-1\rangle \vert 1\rangle_{aux}\}$, the
collective unitary transformation $U_{max}$ can be chosen as
\begin{widetext}
\begin{center}
\begin{eqnarray}
U_{max}=\left(
\begin{array}{cccccccc}
\frac{c_{k}}{c_{0}}              &  0                               & \cdots  & 0                                  & \sqrt{1-(\frac{c_{k}}{c_{0}})^2}   & 0                                 & \cdots  & 0 \\
0                                & \frac{c_{k}}{c_{1}}              & \cdots  & 0                                  &  0                                 & \sqrt{1-(\frac{c_{k}}{c_{1}})^2}  & \cdots  & 0 \\
\vdots                           & \vdots                           & \ddots  & \vdots                             & \vdots                             & \vdots                            & \ddots  & \vdots \\
0                                & 0                                & \cdots  & \frac{c_{k}}{c_{d-1}}              & 0                                  & 0                                 & \cdots  & \sqrt{1-(\frac{c_{k}}{c_{d-1}})^2}\\
\sqrt{1-(\frac{c_{k}}{c_{0}})^2} & 0                                & \cdots  & 0                                  & -\frac{c_{k}}{c_{0}}                & 0                                 & \cdots  & 0 \\
0                                & \sqrt{1-(\frac{c_{k}}{c_{1}})^2} & \cdots  & 0                                  &  0                                 & -\frac{c_{k}}{c_{1}}               & \cdots  & 0\\
\vdots                           & \vdots                           & \ddots  & \vdots                             & \vdots                             & \vdots                            & \ddots  & \vdots\\
0                                & 0                                & \cdots  & \sqrt{1-(\frac{c_{k}}{c_{d-1}})^2} & 0                                  & 0                                 & \cdots  & -\frac{c_{k}}{c_{d-1}}  \\
\end{array}
\right),
\end{eqnarray}
\end{center}
\end{widetext}
i.e.,
\begin{widetext}
\begin{center}
\begin{eqnarray}
U_{max}\vert\varphi\rangle_{a_{n+1}}\vert 0\rangle_{aux}&=&
\sum_{j=0}^{d-1}e^{-\frac{2\pi i }{d}[jr+(j\oplus
s)r']}\beta_{j}c_{j \oplus s}\vert j \oplus s\rangle_{a_{n+1}}
\left(\frac{c_{k}}{c_{j \oplus s}}\vert 0\rangle_{aux}+ \sqrt
{1-(\frac{c_{k}}{c_{j \oplus s}})^2}\vert 1\rangle_{aux}\right).
\end{eqnarray}
\end{center}
\end{widetext}

Charlie measures his auxiliary particle after the unitary
transformation $U_{max}$.  The controlled teleportation succeeds if
the measurement result is $\vert 0\rangle_{aux}$; otherwise the
teleportation fails, and the information of the original state is
disappeared. If the controlled teleportation succeeds, Charlie gets
the state of the particle $a_{n+1}$
\begin{eqnarray}
\vert\varphi'\rangle_{a_{n+1}}&=& \sum_{j'=0}^{d-1}e^{-\frac{2\pi i
}{d}[(j'r'+(j'-s)r]}\beta_{d+j'-s \oplus d}c_{k} 
 \vert j' \rangle_{a_{n+1}} \nonumber\\
&=& c_{k} \sum_{j=0}^{d-1}e^{-\frac{2\pi i }{d}[(j\oplus
s)r'+jr]}\beta_{j}\vert j \oplus s\rangle_{a_{n+1}}.
\end{eqnarray}
Charlie can reconstruct the originally unknown state $\vert \chi
\rangle$ by performing a unitary operation
\begin{eqnarray}
U_{r'+r,d-s}=\sum_{j'=0}^{d-1}e^{\frac{2\pi i }{d}  j' (r+r')}
\left\vert j'\oplus d-s\right\rangle \left\langle j'\right\vert
\end{eqnarray}
on his particle $a_{n+1}$, i.e.,
\begin{eqnarray}
U_{r'+r,d-s}\vert\varphi'\rangle_{a_{n+1}}=A\sum_{j=0}^{d-1}\beta_{j}\vert
j\rangle_{a_{n+1}},
\end{eqnarray}
where $A=c_k e^{-\frac{2\pi i }{d}rs}$ is a whole factor which does
not change the feature of the state.

As discussed in Refs. \cite{sjg,Hsu}, the maximal probability  $P_s$
for extracting the unknown state $\vert \chi\rangle$ with the
fidelity unit from the state $\vert\varphi\rangle_{a_{n+1}}$ is the
square of the minimal coefficient in $c_j$ ($j=0,1,\cdots,d-1$).
That is, the receiver Charlie can recover the unknown state $\vert
\chi\rangle$ with the probability $P_s=|c_k|^2$.

\section{Controlled teleportation of $m$ qudits}

Now, let us generalize this scheme to the case with an unknown
$m$-qudit state. In this time, the agents should first shared $m$
pure entangled states $\vert\Phi\rangle^{\otimes m}$ with the same
way discussed above. Similar to the case with an unknown
single-particle qudit state, the sender (Alice)  performs $m$
generalized Bell-state measurements, and then the controllers
(Bob$_s$) make $X_{d}$ measurements on their particles. The receiver
Charlie first probabilistically extracts the information via a
unitary transformation on his particles and an auxiliary two-level
particle, and then reconstructs the original state by performing
some unitary operations on his particles kept.

In detail, the quantum channel is  a sequence of pure entangled
$(n+2)$-particle  states (the same $m$ quantum systems), i.e.,
\begin{eqnarray}
\vert\Phi'\rangle \equiv \prod_{l=1}^{m}
\left(c_{0}\prod_{k=0}^{n+1}\vert 0\rangle_{a_{k}}
+\cdots+\prod_{k'=0}^{n+1}c_{d-1}\vert
d-1\rangle_{a_{k'}}\right)_{l}.\nonumber\\
\end{eqnarray}
Alice sends the $k$-th ($k=1,2,\cdots,n$) particle $a_{kl}$ in the
$l$-th ($l=1,2,\cdots,m$) pure entangled state to Bob$_{k}$ and the
$(n+1)$-th particle $a_{n+1, l}$ to the receiver Charlie, and she
keeps the first particle $a_{0l}$ in each pure entangled state. Also
all the parties can set up this quantum channel with decoy photons
\cite{decoy,decoy2,dengreport,DSQC}, the same as that discussed
above.

Suppose an arbitrary $m$-qudit state can be described as
\begin{eqnarray}
\left\vert \chi' \right\rangle_{\chi_{1}\chi_{2}\cdots \chi_{m}}=
\sum_{n_1'n_2'\cdots n_m'=0}^{d-1}\beta_{n_1'n_2'\cdots
n_m'}\left\vert n_1'n_2'\cdots n_m' \right\rangle,
\end{eqnarray}
and
\begin{eqnarray}
\sum_{n_1'n_2'\cdots n_m'=0}^{d-1} \vert \beta_{n_1'n_2'\cdots
n_m'}\vert^2=1,
\end{eqnarray}
where $ \chi_{1}, \chi_{2}, \cdots$, and $\chi_{m} $ are the $m$
particles in the originally unknown state $\vert \chi' \rangle$. For
the controlled teleportation, Alice first takes the generalized
Bell-state measurement on the particles $\chi_{l}$ and $a_{0l}$
($l=1,2,\cdots,m$), and then the controllers Bob$_{k}$
($k=1,2,\cdots,n$) perform $X_{d}$ measurements on their particles.
The measurements done by all the controllers Bob$_s$ can be written
as
\begin{eqnarray}
M'=\prod_{l=1}^{m} M_{l},
\end{eqnarray}
where
\begin{eqnarray}
M_{l}=(\langle 0\vert_{x})^{n-t_{1}^{l}-\cdots t_{d-1}^{l}}(\langle
1\vert_{x})^{t_{1}^{l}}\otimes\cdots\otimes(\langle
d-1\vert_{x})^{t_{d-1}^{l}}
\end{eqnarray}
represents the single-particle measurements done by all the
controllers on the particles in the $l$-th pure entangled state
$\vert \Phi \rangle_{l} \equiv (c_{0}\prod_{k=0}^{n+1}\vert
0\rangle_{a_{k}} + \cdots+\prod_{k'=0}^{n+1}c_{d-1}\vert
d-1\rangle_{a_{k'}})_{l}$, and $t_{j}^{l}$ represents the number of
the controllers who obtain the outcomes $\vert j\rangle_{x}$.

The state of composite system composed of particles $\chi_{1}$,
$\chi_{2}$, $\cdots$, $\chi_{m}$ and $a_{kl}$ ($k=0,1,\cdots,n+1$
and $l=1,2,\cdots,m$) can be described as
\begin{widetext}
\begin{center}
\begin{eqnarray}
\vert \chi' \rangle \otimes \vert\Phi'\rangle &=&
\sum_{n_1'n_2'\cdots n_m'=0}^{d-1}\beta_{n_1'n_2'\cdots
n_m'}\left\vert n_1'n_2'\cdots
n_m'\right\rangle_{\chi_{1}\chi_{2}\cdots \chi_{m}} \otimes
\prod_{l=1}^{m} \left(c_{0}\prod_{k=0}^{n+1}\vert 0\rangle_{a_{k}} +
\cdots+\prod_{k'=0}^{n+1}c_{d-1}\vert
d-1\rangle_{a_{k'}}\right)_{l}\nonumber\\
&=& \frac{1}{d^{m/2}} \sum_{r_{1}\cdots r_{m},\atop s_{1}\cdots
s_{m}, j_{1}\cdots j_{m}}^{d-1}\vert
\psi_{r_{1}s_{1}}\rangle_{\chi_1 a_{01}}\otimes \vert
\psi_{r_{2}s_{2}}\rangle_{\chi_2 a_{02}}\otimes \cdots \otimes \vert
\psi_{r_{m}s_{m}}\rangle_{\chi_m a_{0m}} \otimes  e^{-\frac{2\pi i
}{d}(j_{1}r_{1}+ j_{2}r_{2} + \cdots +
j_{m}r_{m})} \nonumber\\
&& \otimes \beta_{j_{1}j_{2} \cdots j_{m}} \otimes c_{j_{1}\oplus
s_1}c_{j_{2}\oplus s_2} \cdots c_{j_{m} \oplus s_m}
\left(\prod_{k_1=1}^{n+1}\vert j_1\oplus s_1\rangle _{k_1} \right)
\left(\prod_{k_2=1}^{n+1}\vert j_2\oplus s_2\rangle _{k_2} \right)
\cdots \left(\prod_{k_m=1}^{n+1}\vert j_m\oplus s_m\rangle _{k_m}
\right).\nonumber \\
\end{eqnarray}
\end{center}
\end{widetext}
That is, after Alice performs $m$ GBS measurements on her $2m$
particles $\chi_la_{0l}$ ($l=1,2,\cdots, m$), the  subsystem
composed of the particles remained collapses to the corresponding
state $\vert\xi\rangle_{a_{11}a_{12}\cdots a_{n+1,m}}$. If the
outcomes of the GBS measurements obtained by Alice are $\vert
\psi_{r_ls_l}\rangle_{\chi_la_{0l}}$ ($l=1,2,\cdots, m$), the state
of the subsystem can be written as (without normalization)
\begin{widetext}
\begin{center}
\begin{eqnarray}
\vert\xi\rangle_{a_{11}a_{12}\cdots a_{n+1,m}}&=& \sum_{j_{1}\cdots
j_{m}=0}^{d-1} e^{-\frac{2\pi i }{d}(j_{1}r_{1}+ j_{2}r_{2} + \cdots
+ j_{m}r_{m})}  \beta_{j_{1}j_{2} \cdots j_{m}} c_{j_{1}\oplus
s_1}c_{j_{2}\oplus s_2} \cdots c_{j_{m} \oplus s_m}\nonumber\\
&& \otimes \left(\prod_{k_1=1}^{n+1}\vert j_1\oplus s_1\rangle
_{k_1} \right) \left(\prod_{k_2=1}^{n+1}\vert j_2\oplus s_2\rangle
_{k_2} \right) \cdots \left(\prod_{k_m=1}^{n+1}\vert j_m\oplus
s_m\rangle _{k_m} \right).
\end{eqnarray}
\end{center}
\end{widetext}

After all the controllers Bob$_s$ take single-particle measurements
on their particles with the MB $X_d$, the state of the particles
$a_{n+1,l}$ ($l=1,2,\cdots,m$) kept by the receiver Charlie becomes
\begin{widetext}
\begin{center}
\begin{eqnarray}
\vert \theta \rangle_{a_{n+1,1}a_{n+1,2}\cdots a_{n+1,m}} &\equiv&
M'\vert\xi\rangle_{a_{11}a_{12}\cdots a_{n+1,m}}\nonumber\\
&=& \sum_{j_{1}\cdots j_{m}=0}^{d-1} e^{-\frac{2\pi i
}{d}\{[j_{1}r_{1}+(j_{1}\oplus s_{1})r''_{1}]+
[j_{2}r_{2}+(j_{2}\oplus s_{2})r''_{2}] + \cdots +
[j_{m}r_{m}+(j_{m}\oplus s_{m})r''_{m}]\}} \otimes \beta_{j_{1}j_{2} \cdots j_{m}}\nonumber\\
&&  \otimes  c_{j_{1}\oplus s_1}c_{j_{2}\oplus s_2} \cdots c_{j_{m}
\oplus s_m}  \otimes \vert j_1\oplus s_1\rangle _{a_{n+1,1}} \vert
j_2\oplus s_2\rangle _{a_{n+1,2}} \cdots \vert j_m\oplus s_m\rangle
_{a_{n+1,m}}.
\end{eqnarray}
\end{center}
\end{widetext}
Here $ r''_{l}=t_{1}^{l}+ 2t_{2}^{l} + \cdots +(d-1) t_{d-1}^{l}$.
To reconstruct the original state probabilistically, Charlie first
performs a unitary transformation on his particles and an auxiliary
particle whose original state is $\vert 0\rangle_{aux}$. In essence,
the auxiliary particle is used to select the useful information from
the unknown state, no matter what the useless information is. That
is, Charlie can use a two-dimension qubit (a two-level quantum
system) for extracting the useful information. One level is used to
map the useful information after a unitary evolution, and the other
is used to map some useless information. Similar to the case of
controlled teleportation of an unknown single qudit, under the basis
$\{\vert f g\cdots h\rangle_{a_{n+1,1}a_{n+1,2}\cdots
a_{n+1,m}}\vert0\rangle_{aux}$; $\vert f g\cdots
h\rangle_{a_{n+1,1}a_{n+1,2}\cdots a_{n+1,m}}\vert1\rangle_{aux}\}$
($f,g,h \in \{0,1$, $\cdots$, $d-1\}$) the unitary evolution ($2d^m
\times 2d^m$ matrix) can be chosen as
\begin{widetext}
\begin{center}
\begin{eqnarray}
 U'_{max}=
 \left(
\begin{array}{cccccccccc}
\frac{(c_{k})^m}{(c_{0})^m}        & \cdots    &  0                    & \cdots  & 0                                     & \sqrt{1-(\frac{c_{k}}{c_{0}})^{2m}}    & \cdots  & 0                      & \cdots  & 0 \\
\vdots                             & \ddots    & \vdots                & \ddots  & \vdots                                & \vdots                                 & \ddots  & \vdots                 & \ddots  & \vdots \\
0                                  & \cdots    & \Gamma_{fg\cdots h}   & \cdots  & 0                                     & 0                                      & \cdots  & \Gamma^+_{fg\cdots h}  & \cdots  & 0 \\
\vdots                             & \vdots    & \vdots                & \ddots  & \vdots                                & \vdots                                 & \vdots  & \vdots                 & \ddots  & \vdots \\
0                                  & \cdots    & 0                     & \cdots  & \frac{(c_{k)^m}}{(c_{d-1})^m}         & 0                                      & \cdots  & 0                      & \cdots  & \sqrt{1-(\frac{c_{k}}{c_{d-1}})^{2m}}\\
\sqrt{1-(\frac{c_{k}}{c_{0}})^{2m}}& \cdots    & 0                     & \cdots  & 0                                     & -\frac{(c_{k})^m}{(c_{0})^m}            & \cdots  & 0                      & \cdots  & 0 \\
\vdots                             & \ddots    & \vdots                & \ddots  & \vdots                                & \vdots                                 & \ddots  & \vdots                 & \ddots  & \vdots \\
0                                  & \cdots    &\Gamma^+_{fg\cdots h}  & \cdots  & 0                                     & 0                                      & \cdots  & -\Gamma_{fg\cdots h}    & \cdots  & 0\\
\vdots                             & \vdots    & \vdots                & \ddots  & \vdots                                & \vdots                                 & \vdots  & \vdots                 & \ddots  & \vdots\\
0                                  & \cdots    & 0                     & \cdots  & \sqrt{1-(\frac{c_{k}}{c_{d-1}})^{2m}} & 0                                      & \cdots  & 0                      & \cdots  &-\frac{(c_{k)^m}}{(c_{d-1})^m}  \\
\end{array}
\right),
\end{eqnarray}
\end{center}
\end{widetext}
where
\begin{eqnarray}
\Gamma_{fg\cdots h}\equiv \frac{(c_{k})^m}{c_{f}c_{g}\cdots c_{h}},
\;\;\;\Gamma^+_{fg\cdots h}\equiv \sqrt{1-(\Gamma_{fg\cdots h})^2}.
\end{eqnarray}
That is, the unitary evolution $U'_{max}$ can transfer the state
$\vert \theta \rangle_{a_{n+1,1}a_{n+1,2}\cdots a_{n+1,m}}$ into the
unknown state $\left\vert \chi'
\right\rangle_{\chi_{1}\chi_{2}\cdots \chi_{m}}$ probabilistically,
i.e.,
\begin{widetext}
\begin{center}
\begin{eqnarray}
U'_{max} \vert \theta \rangle_{a_{n+1,1}a_{n+1,2}\cdots
a_{n+1,m}}\vert 0\rangle_{aux} &=& \sum_{j_{1}\cdots j_{m}}
e^{-\frac{2\pi i }{d}\{[j_{1}r_{1}+(j_{1}\oplus s_{1})r''_{1}]+
[j_{2}r_{2}+(j_{2}\oplus s_{2})r''_{2}] + \cdots +
[j_{m}r_{m}+(j_{m}\oplus s_{m})r''_{m}]\}} \otimes \beta_{j_{1}j_{2} \cdots j_{m}}\nonumber\\
&&  \otimes  c_{j_{1}\oplus s_1}c_{j_{2}\oplus s_2} \cdots c_{j_{m}
\oplus s_m}  \otimes \vert j_1\oplus s_1\rangle _{a_{n+1,1}} \vert
j_2\oplus s_2\rangle _{a_{n+1,2}} \cdots \vert j_m\oplus s_m\rangle
_{a_{n+1,m}}\nonumber\\
&& \left(\frac{c_{k}^{m}}{ c_{j_{1}\oplus s_1}c_{j_{2}\oplus s_2}
\cdots c_{j_{m} \oplus s_m}}\vert
0\rangle_{aux}+\sqrt{1-(\frac{c_{k}^{m}}{ c_{j_{1}\oplus
s_1}c_{j_{2}\oplus s_2} \cdots c_{j_{m} \oplus s_m}})^{2}}\vert
1\rangle_{aux}\right).\nonumber\\ \label{probabilitym}
\end{eqnarray}
\end{center}
\end{widetext}

Same as the case for controlled teleportation of a single qudit,
Charlie perform a measurement on the auxiliary qubit with the MB
$\{\vert 0\rangle, \vert 1\rangle\}$. The controlled teleportation
fails if the measurement result is $\vert 1\rangle_{aux}$;
otherwise, the teleportation succeeds and the particles kept by
Charlie will collapse to the state
\begin{widetext}
\begin{center}
\begin{eqnarray}
\vert \theta' \rangle_{a_{n+1,1}a_{n+1,2}\cdots a_{n+1,m}} &=&
c_{k}^{m} \sum_{j_{1}\cdots j_{m}} e^{-\frac{2\pi i
}{d}\{[j_{1}r_{1}+(j_{1}\oplus s_{1})r''_{1}]+
[j_{2}r_{2}+(j_{2}\oplus s_{2})r''_{2}] + \cdots +
[j_{m}r_{m}+(j_{m}\oplus s_{m})r''_{m}]\}} \otimes \beta_{j_{1}j_{2} \cdots j_{m}}\nonumber\\
&&  \otimes \vert j_1\oplus s_1\rangle _{a_{n+1,1}} \vert j_2\oplus
s_2\rangle _{a_{n+1,2}} \cdots \vert j_m\oplus s_m\rangle
_{a_{n+1,m}}\nonumber\\
&=& \alpha \sum_{j_{1}\cdots j_{m}} e^{-\frac{2\pi i
}{d}(j_{1}r'''_{1}+ j_{2}r'''_{2}+ \cdots + j_{m}r'''_{m})}
\beta_{j_{1}j_{2} \cdots j_{m}} \vert j_1\oplus s_1\rangle
_{a_{n+1,1}} \vert j_2\oplus s_2\rangle _{a_{n+1,2}} \cdots \vert
j_m\oplus s_m\rangle
_{a_{n+1,m}}\nonumber\\
\end{eqnarray}
\end{center}
\end{widetext}
where
\begin{eqnarray}
\alpha &=& c^m_{k}e^{-\frac{2\pi i }{d}\left(s_{1}r''_{1}\oplus
s_{2}r''_{2} \oplus \cdots \oplus s_{m}r''_{m}\right)},\\
r'''_{l} &=& r_l + r''_l.
\end{eqnarray}
Charlie can reconstruct the originally unknown state $\left\vert
\chi' \right\rangle_{\chi_{1}\chi_{2}\cdots \chi_{m}}$ with a
unitary transformation determined by the measurement results
published by Alice and the controllers Bob$_s$ if the controlled
teleportation succeeds. Under the basis under the basis $\{\vert f
g\cdots h\rangle_{a_{n+1,1}a_{n+1,2}\cdots a_{n+1,m}}\}$ ($f,g,h \in
\{0,1$, $\cdots$, $d-1\}$), the unitary transformation is
\begin{eqnarray}
&& U_{r'''_{1}r'''_{2}\cdots r'''_{m},s_{1}s_{2}\cdots s_{m}}=
\sum_{j'_{1}j'_{2}\cdots j'_{m}} e^{\frac{2\pi i
}{d}(j'_{1}r'''_{1}+j'_{2}r'''_{2}+\cdots
+j'_{m}r'''_{m})}   \nonumber\\
&&\;\;\;\;\;\; \vert j'_{1}\rangle  \vert j'_{2}\rangle \cdots \vert
j'_{m}\rangle   \langle j'_{1}\oplus s_1 \vert \langle j'_{2}\oplus
s_2 \vert \cdots \langle j'_{m}\oplus s_m\vert,
\end{eqnarray}
i.e.,
\begin{eqnarray}
U_{r'''_{1}r'''_{2}\cdots r'''_{m},s_{1}s_{2}\cdots s_{m}}\vert
\theta' \rangle=\alpha \left\vert \chi'
\right\rangle_{\chi_{1}\chi_{2}\cdots \chi_{m}}.
\end{eqnarray}

From Eq.(\ref{probabilitym}), one can see the maximal probability
that Charlie can reconstruct the originally unknown state
$\left\vert \chi' \right\rangle_{\chi_{1}\chi_{2}\cdots \chi_{m}}$
with the fidelity unit is $P_{sm}=|c_k|^{2m}$. Here $\vert
c_k\vert^2=min\{\vert c_j\vert^2, j=0,1, \cdots, d-1\}$.

\section{Discussion and summary}

If $c_{j}=1$ for all the $j$ from 0 to $d -1$, the quantum channel
is composed of $m$ maximally entangled $(n+2)$-particle states. The
receiver can reconstruct the unknown state with probability 100\% in
principle if he cooperates with all the controllers, similar to the
case with two-level quantum systems in Ref. \cite{lixhjpb}.
Moreover,  the unitary evolution $U'_{max}$ is the identity matrix
$I_{2d^m\times 2d^m}$ which means doing nothing on the particles
controlled by the receiver and his auxiliary two-level particle. The
receiver can obtain the unknown state with $m$ single-particle
unitary operations on his $m$ particles.

In summary, we have presented a general scheme for
multiparty-controlled teleportation of an arbitrary $m$-qudit state
by using $m$ pure entangled (n+2)-particle quantum systems as the
quantum channel. The sender Alice can share a sequence of pure
entangled states with all the other parties by inserting some decoy
photons randomly in the particle sequences transmitted to the
controllers and the receiver. The receiver can probabilistically
extract the information of the originally unknown state by
performing a general evolution on his particle and an auxiliary
two-level particle.  Charlie can reconstruct the originally unknown
state with $m$ unitary transformations on his particles according to
the measurement results obtained by all the parties. The optimal
probability of successful teleportation is $p=|c_{k}|^{2m}$ which is
just the entanglement of the quantum channel.

\section*{Acknowledgements}
This work was supported by the National Natural Science Foundation
of China under Grant Nos. 10604008 and 10435020,  and Beijing
Education Committee under Grant No. XK100270454.

\end{document}